\documentclass[11pt]{article}

\pdfoutput=1

\usepackage[T1]{fontenc}
\usepackage[utf8]{inputenc}
\usepackage{setspace}
\usepackage{amsmath}
\usepackage{accents}  
\usepackage{amsxtra}  
\usepackage{amsthm}  
\usepackage{amsfonts}  
\usepackage{esint}  
\usepackage{bm}
\usepackage{bbm}
\usepackage{bbold}

\usepackage{gensymb}  
\usepackage{units}

\usepackage{float}
\usepackage{flafter}
\usepackage{afterpage}  
\usepackage{caption}

\usepackage{graphicx}  
\usepackage{array}  

\usepackage[authoryear,round,sort&compress]{natbib}  

\usepackage[bookmarks=false,hypertex,colorlinks=false,citebordercolor={1 1 1},filebordercolor={1 1 1},linkbordercolor={1 1 1},menubordercolor={1 1 1},urlbordercolor={1 1 1},runbordercolor={1 1 1}]{hyperref}

\usepackage{listings}

\makeatletter

\theoremstyle{plain}

\theoremstyle{definition}

\theoremstyle{remark}

\newcommand\Prob{\text{P}}  

\makeatother

\begin{document}

\title{Predicting phenological events using event-history analysis}
\author{Song Cai, James V. Zidek\thanks{Department of Statistics, the
University of British Columbia, 333-6356 Agricultural Road, Vancouver, BC, V6T
1Z2}, Nathaniel Newlands\thanks{Environmental Health, Agriculture and
Agri-Food Canada, 5403 - 1st Avenue S., P.O. Box 3000, Lethbridge, Alberta,
Canada}}

\maketitle

\begin{abstract}

This paper presents an approach to phenology, one based on the use of a method
developed by the authors for event history data. Of specific
interest is the prediction of the so-called ``bloom--date'' of fruit trees in the
agriculture industry and it is this application which we consider, although
the method is much more broadly applicable. Our approach provides sensible
estimate for a parameter that interests phenologists -- $T_{base}$, the
thresholding parameter in the definition of the growing degree days (GDD). Our
analysis supports scientists' empirical finding: the timing of a phenological
event of a prenniel crop is related the cumulative sum of GDDs. Our prediction
of future bloom--dates are quite accurate, but the predictive uncertainty is
high, possibly due to our crude climate model for predicting future
temperature, the time-dependent covariate in our regression model for
phenological events. We found that if we can manage to get accurate prediction
of future temperature, our prediction of bloom--date is more accurate and the
predictive uncertainty is much lower.

\end{abstract}

\section{Introduction}

Phenology, in agricultural science, is the study of periodic plant
developmental stages and their responses to climate (especially to seasonal
and interannual variations in climate) and other physical variables (e.g.
photoperiod). For example, an apple tree, in each of its development cycles,
may go through developmental events from bud-bursting, blooming to fruiting.
It is well known in the agricultural science community that climate variables,
especially daily average temperature, and possibly photoperiod are major
factors that influence the timings of phenological events. But one question remains: how to model their relationship so as to
predict the timings of future phenological events?

Many empirical biological models have been built for representing the
relationship between phenological events and climate variables (see
\citealp{Chuine2000a}, for a comprehensive review). These models are
deterministic and the values of parameters in the models are either determined
experimentally or obtained as point estimates given by least squares to yield
best fits to observed data. The uncertainties associated with these parameter estimates and predictions are often not assessed.

Statistical models have also been applied to phenology. Ordinary least square
(OLS) linear regression is widely used to study the linear association between
timings of phenological events and climate variables. Survival
data analysis techniques, such as the Cox model,  have also been applied (e.g.
\citealp{Gienapp2005}).  However statistical issues arising in phenological
data analysis are quite complicated. Firstly, the phenological events are
irreversible progressive events -- a tree cannot bloom without having gone
through bud-bursting and leafing, and once it blooms, it cannot repeat the
earlier stages in the same development cycle. Secondly, the climate variables
are external time-dependent covariates (i.e. covariates not influenced by the
occurrence of the events of central interest, \citealp{Kalbfleisch2002}). When
time-dependent covariates are present,  OLS linear regression is not
suitable for prediction since the covariates values at future event times
are unknown. The Cox model is also not suitable for prediction because it does
not extrapolate beyond the last observation. Furthermore, the Cox model may be
subject to substantial loss of efficiency due to the strong trends in climate
covariates.

This paper presents an approach to prediction of the bloom--dates of perennial
crops based on time-dependent climate covariates using a regression model
developed by authors for progressive event history data \citep{Cai2010}. This
approach can incorporate all time-dependent covariate information, so there is no
loss of efficiency. Also, prediction is easily formulated in this framework.
Finally this approach can be applied in other areas of application, for example in medical research, to the analysis of survival data with progressive health outcomes.

The report is organized as follows. In Section \ref{sect:data} we describe the data used in our application along with its objectives.  Then in Section \ref{sec:method} we describe the approach used in our analysis. That analysis is presented in Section \ref{sec:analysis}. Finally conclusions are given in Section \ref{sec:conclusions}.

\section{Data and objectives}\label{sect:data}

The data represents the bloom--dates of six high-value perennial
agricultural crops (apricot, cherry, peach, prune, pear, and apple) in
Summerland, British Columbia, Canada, recorded during the 
years between 1937 and 1964 inclusive. 
Each year, the blooming event occurs at most once for each crop. The bloom
date is then recorded as the number of days from the first day of a year to a
``representative'' bloom--date of all the trees in the area in that year. Daily
maximum and minimum temperatures in the same area in the corresponding years are also recorded.

Phenological studies (e.g. \citealp{Murray1989}) suggest that the occurrence
of a phenological event may be mainly influenced by the accumulation of the
so-called growing degree days (GDD),
\begin{equation}
  GDD\left(t \right) = \left\{ \begin{array}{cc}
    \frac{T_{min}\left(t \right)+T_{max}\left(t \right)}{2} - T_{base} &
    \text{if} \enskip \frac{T_{min}\left(t \right)+T_{max}\left(t \right)}{2}
    > T_{base} \\
    0 & \text{otherwise} \end{array} \right., \ \text{for } t > t_0 \;,
    \label{eq:GDDdef}
\end{equation}
where the time $t$ is recorded in days,  $t_0$ is a well-defined temporal 
origin, and $T_{min}\left(t \right)$ (respectively $T_{max}\left(t \right)$) is the daily
minimum (respectively maximum) temperature,  The constant 
$T_{base}$ is a unknown constant threshold 
temperature. The time origin $t_0$ is usually chosen as the
start--date of a development stage \citep{Chuine2000a}. Here, for blooming, we choose $t_0$ as January $1^{st}$. Although this choice may have an impact on
the analysis we do not believe it to be significant. For as can be seen from
(\ref{eq:GDDdef}), the actual start--date is controlled by $T_{base}$. If we
choose some date earlier than that start--date, the daily average
temperatures, $\frac{T_{min}\left(t \right)+T_{max}\left(t \right)}{2}$, in
those earlier days will be smaller than $T_{base}$, and the corresponding GDDs
will be 0. Therefore, that choice will not impact the results of the analysis. For
the bloom--date, the choice of January $1^st$ may be far earlier than the
start--date of the blooming stage, and so its impact on results should be
negligible. 

The results of this paper aim to answer the following questions: (1) In what form of aggregation does the GDD most influence the date of the blooming event. More specifically, is it the cumulative sum of the GDD, a weighted cumulative sum, or something else? (2) What is the value of $T_{base}$? (3) Most importantly, how should the future bloom--date be best predicted the and how should the uncertainty associated with the prediction be best assessed?
From a statistical scientist's perspective, the first question is a model selection
problem, the second, an estimation problem, and the last, a 
prediction problem. 

\section{Methodology} \label{sec:method}

This paper uses a regression method for a single
phenological event developed by authors \citep{Cai2010}. This method is based on a model that uses the observed (discrete) process that represents the state indicator of a phenological event. The blooming event is a single progressive event, i.e. in the development cycle of a plant, once it blooms, the plant stays in
the ``occurred'' state and cannot return to the ``not--occurred'' state. At each
time $t$, we denote the state of the event by an indicator $Y_t$, being
$0$ or $1$ according as the event has ``occurred'' or not. It can be shown
that for such an event, the process of the state indicator $Y_t$ is a Markov
chain. Let $\mathcal{X}_{t}$ be an associated time-dependent covariate vector,
and $\Prob_{t}\equiv\Prob\left(Y_t=1\lvert Y_t=0,\,\mathcal{X}_{t}\right)$,
where $\Prob\left(\cdot\right)$ is a probability set function. Then at each
time point $t$, we can consider a model for the binary event $Y_t$:
\begin{equation}
  g\left(\Prob_{t}\right) = f\left(\mathcal{X}_{t};\,\beta\right) \ ,
  \label{eq:regmodel}
\end{equation}
where $g: \left(0,\,1\right) \rightarrow \left( -\infty,\, \infty \right)$ is
a monotonic link function, and $f: \left( -\infty,\, \infty \right)
\rightarrow \left( -\infty,\, \infty \right)$ is a function of
$\mathcal{X}_{t}$ with parameter vector $\beta$, which encodes the
relationship between $\Prob_{t}$ and $\mathcal{X}_{t}$. Here we
take $g$ to be the logit funcion and restrict $f$ to be a linear function of $\beta$.
One then can derive the probability that a plant blooms at any time point $t$
after the time origin $t_0$. If there are $N$ independent observations, the
likelihood function of the data then can be easily written down accordingly.
Up to this point, the parameter vector $\beta$ can be estimated by the maximum
likelihood (ML) method.

After the model parameters have been estimated, the fitted model can be used to
predict future bloom--dates. For a new year, in which the bloom--date is
unknown, take the time origin $t_0$ as January $1^{st}$ and suppose the
current time is $t_c \ge t_0$, up to which the blooming event has not
occurred. Denote the unknown bloom--date of this new year as $T^*$, with
corresponding state indicator $Y^*_t$ at time $t\ge t_0$. Now, since the bloom
date of a plant usually is related to the associated climate covariates up to
the bloom--date itself, we need to predict the future values of the climate
variables first. For this purpose we fit a ARIMA time series model. Then at any 
time $t\ge t_0$, we denote the covariate vector associated with $Y^*_{t}$ by
$\mathcal{X}^*_{t}$. Because we have observed the value of $\mathcal{X}^*_{t}$
up to the current time $t_c$, we may decompose $\mathcal{X}^*_{t}$ into two parts:
one part $\mathcal{X}^*_{t,\,obs}$ consists of covariates evaluated from time
0 to time $t_c$, which we have observed exactly, and the other part
$\mathcal{X}^*_{t,\,pred}$ consists of predicted covariates values from
$t_c+1$ to $t$, whose predictive distributions are given by the ARIMA model.
Treating the maximum likelihood estimate (MLE) of the parameter vector
$\hat{\beta}$ as if it were the true value of the parameter vector, we can
obtain a ``plug-in'' formula for the predictive probability that the blooming
event occurs at time $t_c+K$, for any $K>0$:
\begin{align}
  &\Prob_{\hat{\beta}}\left( T^*=t_c+K \left |
    \mathcal{X}^*_{t_c+K,\,obs} \right. \right)
    \notag \\
    =&
    \int
    \Prob_{\hat{\beta}}\left( T^*=t_c+K \left|
    \mathcal{X}^*_{t_c+K,\,obs}, \; \mathcal{X}^*_{t_c+K,\,pred}
    \right. \right)
    d \Prob \left(\mathcal{X}^*_{t_c+K,\,pred} \right)
    \notag \\
    =&
    \int g^{-1}\left(f \left(\mathcal{X}^*_{t_c+K}; \, \hat{\beta}\right)\right)
    \prod_{s=1}^{K-1}
    \left( 1-g^{-1}\left(f
    \left(\mathcal{X}^*_{t_c+s}; \, \hat{\beta}\right) \right)\right)
    d \Prob \left(\mathcal{X}^*_{t_c+K,\,pred} \right)
    \ .
    \label{eq:preddistplugin}
\end{align}
Generate a sample of large size $L$ (e.g. thousands or more) from the
predictive
distribution of $\mathcal{X}^*_{t_c+K,\,pred}$, and denote the sample points
as $\mathcal{X}^*_{t_c+K,\,pred}\left(l \right)$ ($l=1,\,\cdots,\,L$). The
above predictive probability then can be approximated by Monte Carlo (MC)
integration, 
\begin{equation}
  \Prob_{\hat{\beta}} \left( T^*=t_c+K \lvert \mathcal{X}^*_{t_c+K,\, obs}\right)
  \approx \frac{1}{L}\sum_{l=1}^L \Prob_{\hat{\beta}}\left( T^*=t_c+K \left|
    \mathcal{X}^*_{t_c+K,\,obs}, \; \mathcal{X}^*_{t_c+K,\,pred}\left(l
    \right) \right. \right) \ .
    \label{eq:predictivedMC}
\end{equation}
Note that in this ``plug-in'' approach, the uncertainties associated with the unknown parameters are not taken into account. However one may use a re-sampling method such as the bootstrap to assess their effect.

\citet{Cai2010} also provides a regression model for multiple phenological
events, which will not be needed here.

\section{Analysis}\label{sec:analysis}

In this section, we apply the method described above to the bloom--dates of the
six different crops separately, and present the results of the analysis.

\subsection{Assumptions}

For each crop, the bloom--date over years is a time series. However, sample
auto-correlations suggest that the auto-correlations of the bloom--dates over
years are negligible for all six crops. We therefore assume that for each
crop, the bloom--dates of different years are independent realizations from the
same population.

\subsection{The relationship between bloom--dates and GDD}

As mentioned above, scientists believe that the bloom--dates are related
to the accumulation of GDD. In particular, empirical results suggest 
that it is the AGDD, the cumulative sum of GDD starting from the time
origin $t_0$, that most influences the timing of the blooming event
(e.g. \citealt{Chuine2000a}, \citealt{Murray1989}):
\begin{equation}
  \text{AGDD} \left(t \right) = \sum_{k=t_0}^{t} \text{GDD}\left(k \right) \ ,
  \label{eq:AGDD}
\end{equation}
where $t_0$ is time origin and $t\ge t_0$ is recorded in days, the same time scale as that of the GDD.

Here we seek by statistical means, the form of GDD aggregation that best
models our data. For example, one might conjecture that 
the GDD evaluated at times near the bloom--date 
would predict the blooming
event better than those in the past. For this model, using a weighted sum of GDDs with weights increasing  over time as a covariate may plausibly yield a better model fit than using AGDD, an unweighted sum of GDDs over time, as a covariate. To investigate this conjecture, we fitted the regression model described in Section \ref{sec:method} to the data with bloom--date as the response,
and we consider the following alternative ways of incorporating GDD as a covariate
$\mathcal{X}_{t}$:
\begin{description}

  \item [Model AGDD] Take $f \left( \mathcal{X}_{t}; \; \beta \right)$ in
    Equation (\ref{eq:regmodel}) as a linear function of AGDD evaluated at the
    current time $t$:
    \begin{equation}
      f \left( \mathcal{X}_{t}; \; \beta \right) =
      a + b \text{AGDD}\left(t\right)
      = a + b\sum_{k=t_0}^t \text{GDD}\left(k\right) \ ,
    \end{equation}
    where $\mathcal{X}_{t} = \text{AGDD}$ and  the subscript $T$ stands
    for the transpose of a vector or matrix.

  \item [Model ExpSmooth] Take $f \left( \mathcal{X}_{t}; \; \beta \right)$ as
    a linear function of a weighted sum of GDD from the time origin $t_0$ to
    the current time $t$:
    \begin{equation}
      f \left( \mathcal{X}_{t}; \; \beta \right) = a + 
      b\sum_{k=0}^{t-t_0} \left(1 - \gamma\right)^k \text{GDD}\left(t-k\right)
      \ ,
        \label{eq:3expsmooth}
    \end{equation}
    where $\mathcal{X}_{t} = \sum_{k=0}^{t-t_0} \left(1 - \gamma\right)^k
    \text{GDD}\left(t-k\right)$, $\beta=\left(a, \;b, \;\gamma,
    \;T_{base}\right)^T$, and $0\le\gamma\le1$. We call this model
    ``ExpSmooth'' because the weighted average term is similar to the
    exponential smoothing used in time series \citep{Chatfield2004}.

  \item [Model GDD] Take $f \left( \mathcal{X}_{t}; \; \beta \right)$ as a linear function of GDD evaluated
    at the current time $t$:
    \begin{equation}
      f \left( \mathcal{X}_{t}; \; \beta \right) =a + b
      \text{GDD}\left(t\right) \ ,
    \end{equation}
    where $\mathcal{X}_{t} = \text{GDD}\left(t\right)$ and $\beta=\left(a,
    \;b, \;T_{base}\right)^T$.

  \item [Model 5Days] Take $f \left( \mathcal{X}_{t}; \; \beta \right)$ as a linear function of the GDD evaluated at the 5
    most recent days:
    \begin{equation}
      f \left( \mathcal{X}_{t}; \; \beta \right) =a + \sum_{k=1}^5 b_k
      \text{GDD}\left(t-k+1\right) \ ,
    \end{equation}
    where $\mathcal{X}_{t} = \big(\text{GDD}\left(t\right),
    \;\text{GDD}\left(t-1\right), \;\text{GDD}\left(t-2\right),
    \;\text{GDD}\left(t-3\right), \;\text{GDD}\left(t-4\right) \big)^T$ and
    $\beta=\left(a, \;b_1, \;b_2, \;b_3, \;b_4, \;b_5, \;T_{base}\right)^T$

\end{description}
Note that in each of the above models, $T_{base}$ is a parameter
included in the expression of GDD.

Model AGDD incorporates the empirical results referred to above, 
and it serves as a basis of
comparison. Model GDD is used to assess whether the probability of blooming is influenced mainly by the GDD evaluated at the current time. Model 5Days tests
the theory that the GDD evaluated at each of many time points prior to the
current time might be important predictors,  
each having a different effect on $P_{t}$. In the latter model, we give each GDD evaluated at several days
prior to and at the current day a different regression coefficient. However
we consider only GDD evaluated at the five most recent days, because given 28
years of bloom--dates, we won't be able to get good estimates of model
parameters, if the number of parameters is too large.

We found the most promising model to be Model ExpSmooth. This model, $f \left(
\mathcal{X}_{t}; \; \beta \right)$ is a linear function of the weighted sum of
GDD evaluated from the time origin $t_0$ to the current time $t$. For a fixed
$\gamma$ ($0\le\gamma\le1$), $\left(1-\gamma\right)^k$, the weight on
the GDD at lag $k$, the number of days prior to the current date, decays as $k$
increases. This reflects the idea that the GDD evaluated at recent time points
contribute more to the probability of occurrence of the blooming event at the
current time than do the GDD evaluated at time points long before the current
time $t$.

\begin{figure}[!ht]
\begin{center}
    \includegraphics[scale=0.6]{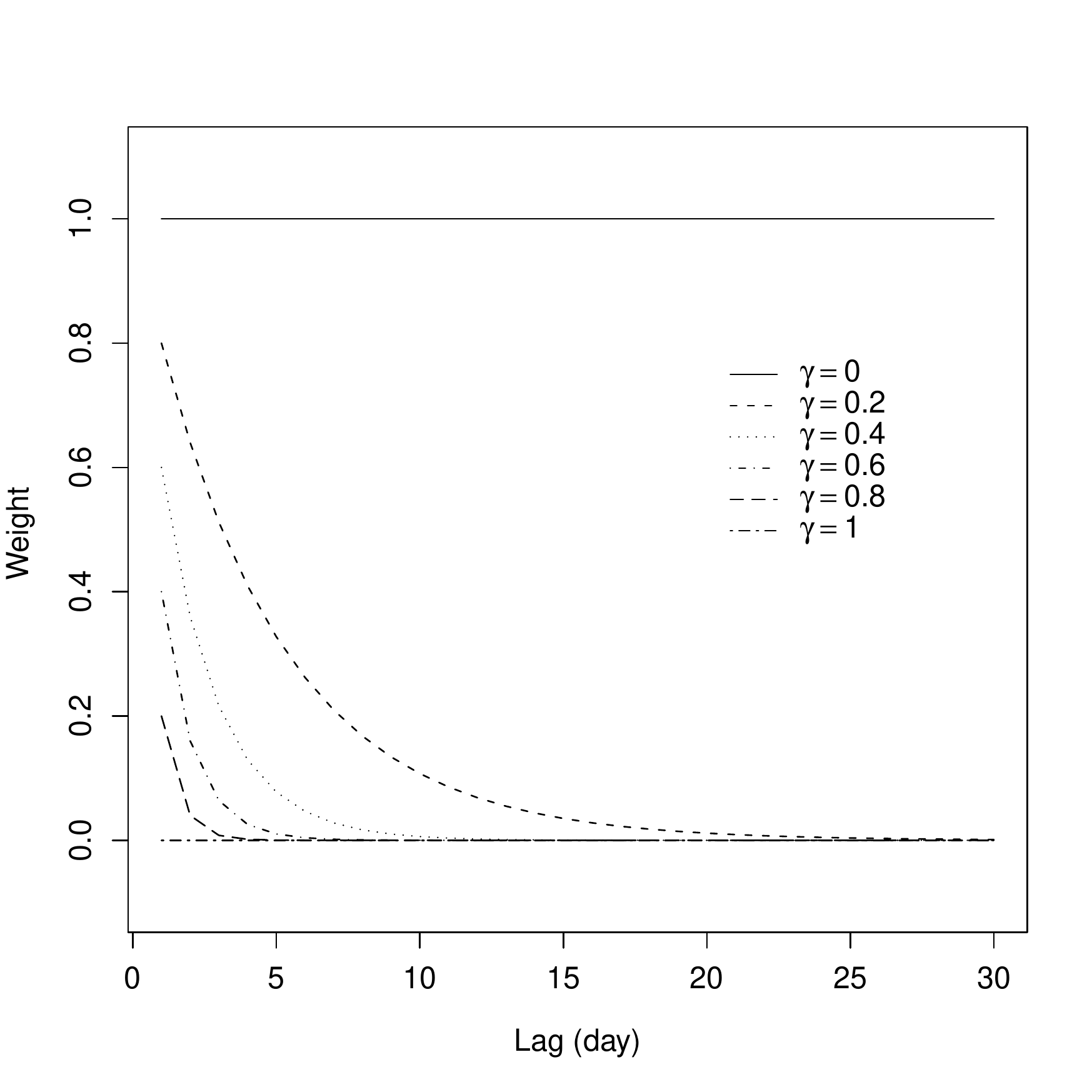}
    \caption{The actual weights in the weighted sum in Model ExpSmooth for
    different $\gamma$ parameter values. The weight decays when the lag
    (number of days prior to the current date) increases. A larger $\gamma$
    corresponds to a faster speed of decaying.}
    \label{fig:3expsmooth}
\end{center}
\end{figure}

In Model ExpSmooth, the value of $\gamma$ controls the speed of with which the  weight decays. Figure \ref{fig:3expsmooth} shows how the weight decays when the lag
increases for different values of $\gamma$.
When $\gamma$ becomes larger, the weight decays faster. In the extreme case of
$\gamma=1$, the weighted sum is just GDD evaluated at the current time and so
Model ExpSmooth becomes Model GDD. If $\gamma$ becomes smaller, the weight
decays slower. In the extreme case of $\gamma=0$, i.e. no decay, Model
ExpSmooth becomes Model AGDD. In other words, Model GDD and AGDD are only
special cases of Model ExpSmooth. The value of $\gamma$, however, is not known
in advance. Thus, we treat it as a model parameter, and estimate it using the
maximum likelihood estimator (MLE).

\begin{table}[!ht]
\begin{center}
  \caption{BICs of the fitted models}
  \label{tab:BIC}
\begin{tabular}{|l|cccc|}
  \hline
   & AGDD & ExpSmooth & GDD  & 5Days \\
  \hline
  Apricot &148.70 &147.53 &249.65 &239.01 \\
  Cherry  &163.69 &157.41 &255.72 &243.73 \\
  Peach   &146.68 &146.49 &255.06 &255.25\\
  Prune   &142.79 &141.91 &224.14 &232.26\\
  Pear    &132.57 &133.14 &231.19 &224.63\\
  Apple   &126.31 &128.79 &265.11 &254.44\\
  \hline
\end{tabular}
\end{center}
\end{table}

For every crop, we fit all the above models to the data, and compare the
Bayesian information criterion (BIC) for each with the results in 
Table \ref{tab:BIC}. Clearly, for all crops Model AGDD and
ExpSmooth are essentially equivalent and both are 
much better than the other two models. The
estimated smoothing parameter, $\hat\gamma$, in Model ExpSmooth for different crops
are shown in Table \ref{tab:gamma}.
\begin{table}[!ht]
\begin{center}
  \caption{Estimated smoothing parameter $\gamma$ of Model ExpSmooth for
  different crops}
  \label{tab:gamma}
\begin{tabular}{|cccccc|}
  \hline
   Apricot & Cherry & Peach  & Prune & Pear & Apple \\
  \hline
   0.014 & 0.020 & 0.0083 & 0.016 & 0.023 & 0.0036 \\
  \hline
\end{tabular}
\end{center}
\end{table}
All $\hat\gamma$'s are very small, which suggests that the
weights in the fitted Model ExpSmooth's decay very slow, and therefore that the
fitted Model ExpSmooth resembles the fitted Model AGDD for our data.
This statistical result supports scientists' experimental result: the
accumulation of GDD is roughly in the form of a sum with equal weights, in other words the AGDD.

Although the quality of the models, ExpSmooth and Model AGDD, seem  roughly equivalent, we study only Model AGDD for the following reasons: (1) it has been the traditional choice; (2) it is more parsimonious, having  one less parameter, making it preferable for the small samples we need to deal with.

\subsection{Estimating model parameters} \label{subsec:est}

The estimates of parameters in Model AGDD in the above discussion were obtained by maximum likelihood (ML) and these are shown in Table \ref{tab:estPar}.
\begin{table}[!ht]
\begin{center}
  \caption{Estimated parameters of Model AGDD} \label{tab:estPar}
\begin{tabular}{|l|ccc|}
  \hline
  Model & $\hat{a}$ & $\hat{b}$ & $\hat{T}_{base}$ \\
  \hline
  Apricot & -13.49 & 0.061 & 2.65 \\
  Cherry & -11.72 & 0.043 & 3.35 \\
  Peach & -19.67 & 0.043 & 0.38 \\
  Prune & -18.23 & 0.057 & 2.80 \\
  Pear & -22.27 & 0.07 & 2.97 \\
  Apple & -26.77 & 0.07 & 2.82 \\
   \hline
\end{tabular}
\end{center}
\end{table}
Are these estimated parameters close to the true values of the
parameters? \citet{Wald1949} gave famous sufficient conditions that would ensure that at least as the sample size approaches infinity, the ML estimates 
converge to their true counterparts, a property called consistency. These conditions in turn lead to others that ensure not only consistency but as well, other desirable properties,  namely an approximately normal distribution and asymptotic efficiency. However, one of those conditions requires that the likelihood function be a continuous function of parameters while the definition
of the GDD (\ref{eq:GDDdef}) ensures that the likelihood function for Model AGDD is not a continuous function of $T_{base}$. Thus we cannot apply Wald's theory and we use a different approach to assess estimator quality, namely simulation.

In the simulation study, we generate data as follows. First, we get one year
of long term averaged daily average temperature series by taking the average of the daily average temperature from year 1916 to 2005 in
the Okanagan region of British Columbia for each day of a year.
We then add a noise process to this long term averaged series. This noise
process is generated from an $ARMA(3,\,1)$ model:
\begin{equation}
  X_t= 1.83X_{t-1} - 0.96X_{t-2} + 0.12X_{t-3} + Z_t -0.96Z_{t-1}
  \label{eq:arima}
\end{equation}
where the white noise $Z$ has a normal distribution with mean 0 and variance
5.253 for any $t$. This ARMA model is fitted to the same daily average
temperature used for extracting long term averaged series above. Now we get
one year of simulated daily average temperature data. We calculate the GDD of the generated temperature data with parameter
$T_{base}=3.5$. Now starting from day 1, we generate
a random number $Y_1$ from a Bernoulli distribution $Ber\left(p\right)$ with
parameter $p=logit^{-1}\left( -13 + 0.04\sum_{k=1}^1\text{GDD}_{k}\right)$. If $Y_1=0$,
we generate
\begin{equation}
  Y_2 \sim Ber\Big(logit^{-1} \big( -13 + 0.04\sum_{k=1}^2X_{k} \big)
  \Big) \ .
\end{equation}
Again, as long as $Y_2=0$, we will generate $Y_3$ similarly, and so on 
 until we get a 1 at time $t$. This $t$ is the simulated bloom--date. Using this procedure, we
generate one year of GDDs and a bloom--date for that year, as one year of
data.

Now we generate 30 years of data as one sample (i.e. a sample of size 30), and
we generate 1000 such samples. For each sample $i$ ($i=1,\,\cdots,\,1000$), we
apply Model AGDD, and calculate the MLEs of the model parameters: $\hat{a}_i$,
$\hat{b}_i$, and ${\hat{T}_{base}}_i$. For each parameter, say
$a$, we calculate the estimated mean of the MLE,
\begin{equation}
  \bar{\hat{a}} = \frac{1}{1000}\sum_{i=1}^{1000} \hat{a}_i \ ,
\end{equation}
the estimated variance of the MLE,
\begin{equation}
  \frac{1}{1000-1}\sum_{i=1}^{1000} \left(\hat{a}_i-\bar{\hat{a}} \right)^2 \ ,
\end{equation}
and the standard error of the mean of the MLE,
\begin{equation}
  \sqrt{\frac{
  \frac{1}{1000-1}\sum_{i=1}^{1000} \left(\hat{a}_i-\bar{\hat{a}} \right)^2
  }{1000}} \ ,
\end{equation}
which characterize how well the estimated mean approximate the true mean of the
MLE.

We repeat the above procedure for sample sizes $S$ of 80, 150 and 400. If
the MLEs were consistent, we would be able to see as the sample size becomes larger, that for each parameter the estimated mean comes closer to the true
value of the parameter, while the estimated variance becomes smaller. Table
\ref{tab:consMean} shows that when the sample size increases, the estimated
means of the MLEs of $a$ and $b$ become closer to the true parameters values
$a=-13$ and $b=0.04$. When the sample size reaches 400, the estimated means
are basically the true values. For parameter $T_{base}$, the estimated means
using different sample sizes are all fairly close to the true value of
$T_{base}=3.5$. Standard errors of the means (Table \ref{tab:consErr}) show
that these estimated means are reliable. On the other hand, when the sample
size increases, the estimated variances (Table \ref{tab:consVar}) of all
parameters become smaller. These facts suggest that in Model AGDD, the MLEs of
all parameters might be consistent.

\begin{table}[ht]
\begin{center}
  \caption{Estimated means of the MLEs. When the sample sizes increases, the
  estimated means become closer to the true parameter values of $a=-13$,
  $b=0.04$ and $T_{base}=3.5$}
  \label{tab:consMean}
\begin{tabular}{ccccc}
  \hline
  & $S=30$ & $S=80$ & $S=150$ & $S=400$ \\
  \hline
  $\hat{a}$ & -13.82 & -13.23 & -13.20 & -13.07 \\ 
  $\hat{b}$ & 0.043 & 0.041 & 0.041 & 0.040 \\ 
  $\hat{T}_{base}$ ($\degree \text{C}$) & 3.50 & 3.50 & 3.48 & 3.51 \\ 
  \hline
\end{tabular}
\end{center}
\end{table}

\begin{table}[ht]
\begin{center}
  \caption{Standard errors of the means the MLEs. Small standard errors imply
  that the estimated means of MLEs are reliable.}
  \label{tab:consErr}
\begin{tabular}{ccccc}
  \hline
  & $S=30$ & $S=80$ & $S=150$ & $S=400$ \\ 
  \hline
  $\hat{a}$ & 0.066 & 0.035 & 0.026 & 0.015 \\ 
  $\hat{b}$ & 0.0002 & 0.0001 & 0.0001 & 0.0001 \\ 
  $\hat{T}_{base}$ ($\degree \text{C}$) & 0.027 & 0.015 & 0.010 & 0.0065 \\ 
  \hline
\end{tabular}
\end{center}
\end{table}

\begin{table}[ht]
\begin{center}
  \caption{Estimated variances of the MLEs. When the sample size increases,
  the estimated variances become smaller.}
  \label{tab:consVar}
\begin{tabular}{ccccc}
  \hline
  & $S=30$ & $S=80$ & $S=150$ & $S=400$ \\ 
  \hline
  $\hat{a}$ & 4.31 & 1.20 & 0.66 & 0.22 \\ 
  $\hat{b}$ & 0.0001 & 0.0000 & 0.0000 & 0.0000 \\ 
  $\hat{T}_{base}$ ($\degree \text{C}$) & 0.70 & 0.21 & 0.11 & 0.04 \\ 
  \hline
\end{tabular}
\end{center}
\end{table}


\subsection{Assessing the uncertainty of the MLEs}

As noted above, we cannot use standard asymptotic results (e.g. in \citealp{Cox1979}) to find large sample approximations to the standard errors of parameter estimators, forcing us to use an alternative approach. The one we choose, the bootstrap \citep{Efron1994boot} if valid would allow us to not only estimate the standard error of the MLEs but as well to find  quantile based confidence intervals for the model parameters.
However we know of know general theory that would imply that validity in this particular application, leading us to again resort to simulation to explore this issue.

Using the simulated data seen in section \ref{subsec:est},
for each different sample size $S$, we estimate the true variances of the
MLEs by the sample variances of the MLEs obtained using the 1000 samples. The
standard deviation of the MLEs is then estimated by the square root of these
sample variances. The results are shown in the ``Sim.'' fields in Table
\ref{tab:bootEstSd}. We can then see how the bootstrap estimates the
standard deviations of the MLEs compare with the corresponding estimates obtained from the simulated data. 

To get these results, we randomly chose for each different sample size, one sample from the 1000 simulated samples and then 1000 bootstrap samples from this one sample of response and predictor pairs. For each such bootstrap sample, we calculated the MLEs of the parameters. For each parameter, we then took the
square root of the sample variance of the MLEs obtained from the 1000
bootstrap samples,  as the bootstrap estimate of the standard error of the
MLE for that parameter. The results are shown in the ``Boot.'' fields in Table
\ref{tab:bootEstSd}. We can see that for each parameter, when the sample size
becomes large, the bootstrap estimates and the estimates obtained using the
simulated data both become small. The bootstrap estimates are always larger
than the estimates obtained from the simulated data, but when the sample size
gets large, the difference between them becomes small. In fact, for a sample size
of 400, the two estimates are fairly close. This may suggest the bootstrap estimates do converge to the true standard deviations of the MLEs,
although the rate of convergence seems low.

\begin{table}[ht]
\begin{center}
  \caption{Comparison of bootstrap estimates of the standard deviations of the
  MLEs and the estimated standard deviations using simulated data. ``Boot.''
  stands for the bootstrap estimates; ``Sim.'' stands for the estimates
  obtained using simulated data. As the sample size increases, the estimated
  standard deviations calculated using the two different approaches become
  smaller and also closer.}
  \label{tab:bootEstSd}
\begin{tabular}{|l|cc|cc|cc|cc|}
  \hline
  & \multicolumn{2}{c|}{$S=30$} & \multicolumn{2}{c|}{$S=80$} 
    & \multicolumn{2}{c|}{$S=150$} & \multicolumn{2}{c|}{$S=400$} \\
  \hline
    & Boot. & Sim. & Boot. & Sim. & Boot. & Sim. & Boot. & Sim. \\
  \hline
  $\hat{a}$ & 2.23 & 2.08 & 1.55 & 1.10 & 0.71 & 0.81 & 0.54 & 0.46 \\ 
  $\hat{b}$ & 0.0102 & 0.0076 & 0.0050 & 0.0041 & 0.0034 & 0.0031 & 0.0017 &
  0.0018 \\ 
  $\hat{T}_{base} ($\degree \text{C}$)$ & 1.36 & 0.84 & 0.49 & 0.46 & 0.27 & 0.33 & 0.25 & 0.21 \\ 
   \hline
\end{tabular}
\end{center}
\end{table}

We also need to obtain 95\% confidence intervals for the model parameters.
Using the MLEs obtained from the simulated data, we can get quantile-based
confidence intervals for the model parameters. We also can calculate
quantile-based bootstrap confidence intervals using MLEs obtained from the
bootstrap samples of one simulated sample. The results are shown in Table
\ref{tab:bootEstCI}. We see that for each parameter, the lengths of confidence
intervals obtained by the two approaches are roughly the same, and as sample
size gets larger, they both become smaller. However, the confidence intervals
obtained using the two different approaches do not always agree -- the
bootstrap intervals seem to always have a bias. Fortunately, when the sample
size is large, the difference between the two kinds of intervals is pretty
small -- it is alway smaller than 1/20 of the length of the confidence
interval obtained from the simulated data when sample size is 400. We tried a
bias corrected version of quantile based bootstrap confidence interval (``BC''
method in \citealt{Efron1986}), but the results are even slightly worse than
this raw version. Overall, although a small bias may exist, use the quantile-based
bootstrap confidence interval seems reasonable in our application.

\begin{table}[ht]
\begin{center}
  \caption{Comparison of quantile-based 95\% confidence intervals based on
  bootstrap and simulated data. ``Boot.'' stands for the bootstrap estimates;
  ``Sim.'' stands for the estimates obtained using the simulated data. As the
  sample size increases, the confidence intervals calculated using the two
  different approaches both become smaller, but they do not always agree very
  well.}
  \label{tab:bootEstCI}
  \small
\begin{tabular}{|l|c|c|c|c|}
  \hline
  & $S=30$ & $S=80$ & $S=150$ & $S=400$ \\
  \hline
  $a$ (Boot.) & (-17.44, -10.25) & (-17.94, -12.54) & (-13.96, -11.47) &
    (-14.31, -12.51) \\ 
  $a$ (Sim.) & (-17.60, -10.48) & (-15.16, -11.38) & (-14.63, -11.75) &
    (-13.82, -12.20) \\ 
  \hline
  $b$ (Boot.) & (0.034, 0.066) & (0.036, 0.054) & (0.038, 0.050) & (0.036,
    0.042) \\ 
  $b$ (Sim.) & (0.033, 0.061) & (0.035, 0.050) & (0.036, 0.047) & (0.038,
    0.044) \\ 
  \hline
  $T_{base} ($\degree \text{C}$)$ (Boot.) & (2.45, 5.43) & (2.02, 3.79) & (3.64, 4.60) & (2.82,
    3.58) \\ 
  $T_{base} ($\degree \text{C}$)$ (Sim.) & (2.14, 5.13) & (2.72, 4.39) & (2.95, 4.12) & (3.16,
    3.90) \\ 
   \hline
\end{tabular}
\end{center}
\end{table}

Table \ref{tab:bootEstRange} shows the observed range (minimum value to
maximum value) of the bootstrap MLEs. We see that for each parameter and all
the four choices of the sample sizes $S$, this range covers and is much larger
than the 95\% confidence interval obtained using the simulated data. Without
knowing the actual coverage probability, this range cannot be directly used as
a confidence interval. However, the usefulness of it is that if this range
does not contain a value, say $\theta_0$, then we get stronger evidence of
saying that the parameter value is not $\theta_0$ than the possibly biased
95\% bootstrap confidence interval not containing $\theta_0$.
\begin{table}[ht]
\begin{center}
  \caption{Observed ranges of the bootstrap MLEs. These ranges always contain
  the quantile-based 95\% confidence intervals based on the simulated data.}
  \label{tab:bootEstRange}
  \small
\begin{tabular}{|l|c|c|c|c|}
  \hline
  & $S=30$ & $S=80$ & $S=150$ & $S=400$ \\
  \hline
  $\hat{a}$ & (-36.65, -6.43) & (-21.67, -8.55) & (-15.50, -8.98) & (-15.17,
    -7.68) \\ 
  $\hat{b}$ & (0.0055, 0.1527) & (0.0290, 0.0652) & (0.0324, 0.0575) &
    (0.0342, 0.0453) \\ 
  $\hat{T}_{base} ($\degree \text{C}$)$ & (-19.00, 11.44) & (0.90, 6.71) & (3.11, 6.52) & (2.51,
    7.09) \\ 
   \hline
\end{tabular}
\end{center}
\end{table}

The bootstrap estimates of quantile-based 95\% confidence intervals of the
MLEs of Model AGDD for all crops are shown in Table \ref{tab:bootCI}. Given
the data, we are interested in knowing whether the regression coefficients $a$
and $b$ are significantly different from 0. Since none of the 95\% bootstrap
confidence intervals of $a$ and $b$ contains 0, we have a strong evidence that
neither $a$ and $b$ are  0 for all crops. The observed ranges of the bootstrap
MLEs (Table \ref{tab:bootRange}) also support this conclusion -- they do not
cover 0.

\begin{table}[ht]
  \caption{Quantile-based 95\% bootstrap confidence intervals for the model
  parameters}
  \label{tab:bootCI}
\begin{center}
\begin{tabular}{lccc}
  \hline
  & $a$ & $b$ & $T_{base}$ \\ 
  \hline
  Apricot & (-22.43, -12.07) & (0.051, 0.096) & (0.95, 4.00) \\ 
  Cherry & (-21.18, -10.72) & (0.030, 0.095) & (1.01, 5.15) \\ 
  Peach & (-31.69, -16.37) & (0.030, 0.065) & (-2.51, 1.53) \\ 
  Prune & (-31.04, -14.39) & (0.046, 0.093) & (0.18, 4.70) \\ 
  Pear & (-37.95, -16.62) & (0.055, 0.122) & (1.93, 3.81) \\ 
  Apple & (-39.54, -14.66) & (0.060, 0.111) & (1.84, 6.95) \\ 
   \hline
\end{tabular}
\end{center}
\end{table}

\begin{table}[ht]
  \caption{Observed ranges of the bootstrap MLEs}
  \label{tab:bootRange}
\begin{center}
\begin{tabular}{lcc}
  \hline
  & $\hat{a}$ & $\hat{b}$ \\ 
  \hline
  Apricot & (-31.46, -9.49) & (0.029, 0.142) \\ 
  Cherry & (-36.13, -7.56) & (0.015, 0.150) \\ 
  Peach & (-40.64, -6.86) & (0.0090, 0.0871) \\ 
  Prune & (-40.48, -7.48) & (0.018, 0.178) \\ 
  Pear & (-62.80, -6.85) & (0.023, 0.170) \\ 
  Apple & (-50.27, -8.90) & (0.034, 0.169) \\
   \hline
\end{tabular}
\end{center}
\end{table}

\subsection{Prediction}

To use Model AGDD to predict the future bloom--dates of a crop, we need to
first predict the future daily average temperature $\left(
T_{min}\left(t\right) + T_{max}\left(t \right)\right)/2$. Here, we use the
ARIMA model (\ref{eq:arima}) to generate future temperatures. The parameters in
the ARMIA model are estimated by fitting the model to the seasonality-removed
daily average temperature series from year 1916 to 2005 in the Okanagan region
of British Columbia. The order of this ARIMA model is selected by comparing
the BICs of fitted ARIMA$\left(p,\,d,\,q\right)$ models with different orders:
$p$ and $q$ range from 0 to 6, and $d$ ranges from 0 to 4.

\begin{figure}[!ht]
\begin{center}
    \includegraphics[scale=0.6]{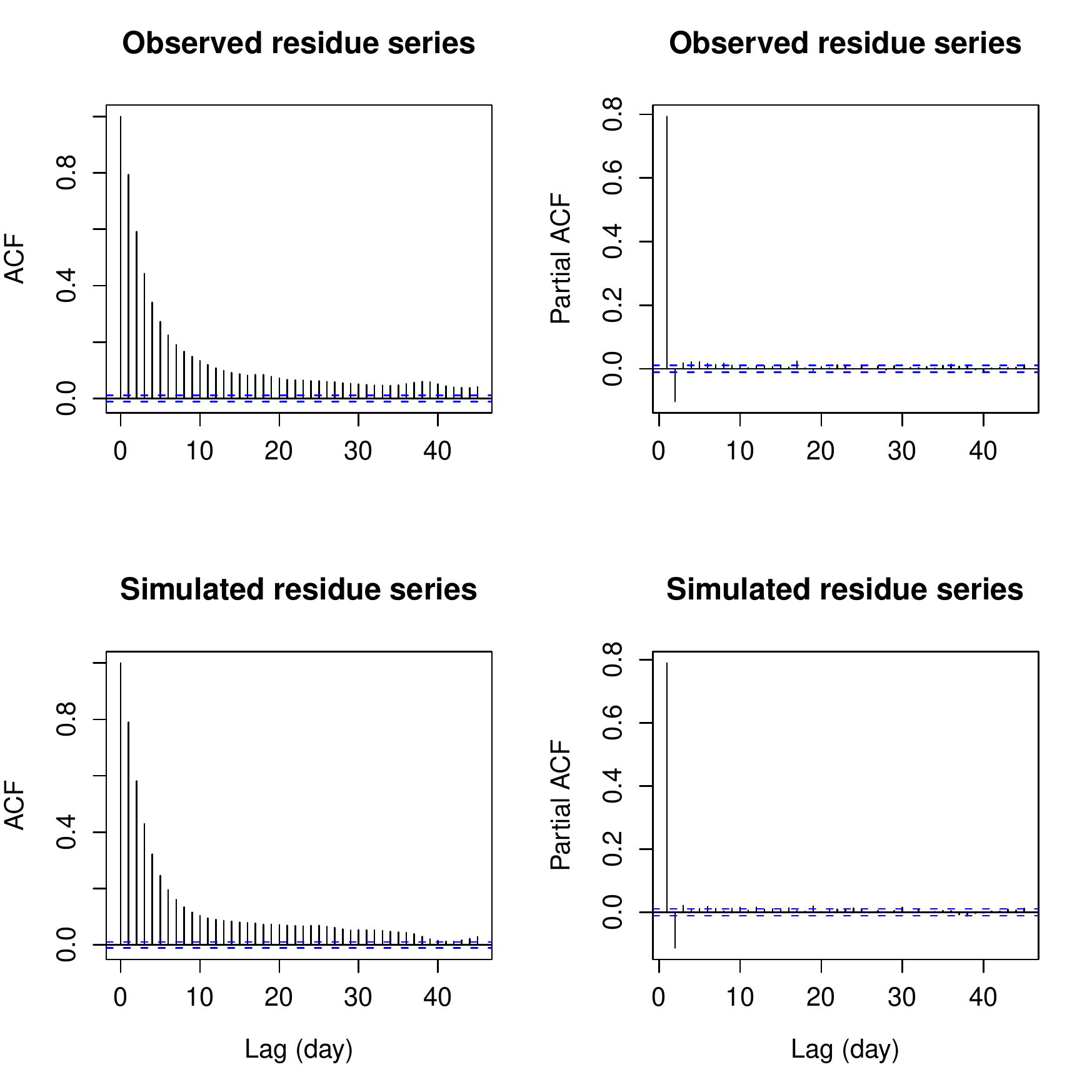}
    \caption{The sample ACF and PACF plots of the observed seasonality-removed
    daily average temperature series and simulated seasonality-removed daily
    average temperature series. The simulated seasonality-removed series have
    similar sample ACF and PACF as the observed seasonality-removed series.}
    \label{fig:acfpacf}
\end{center}
\end{figure}

\begin{figure}[!ht]
\begin{center}
    \includegraphics[scale=0.6]{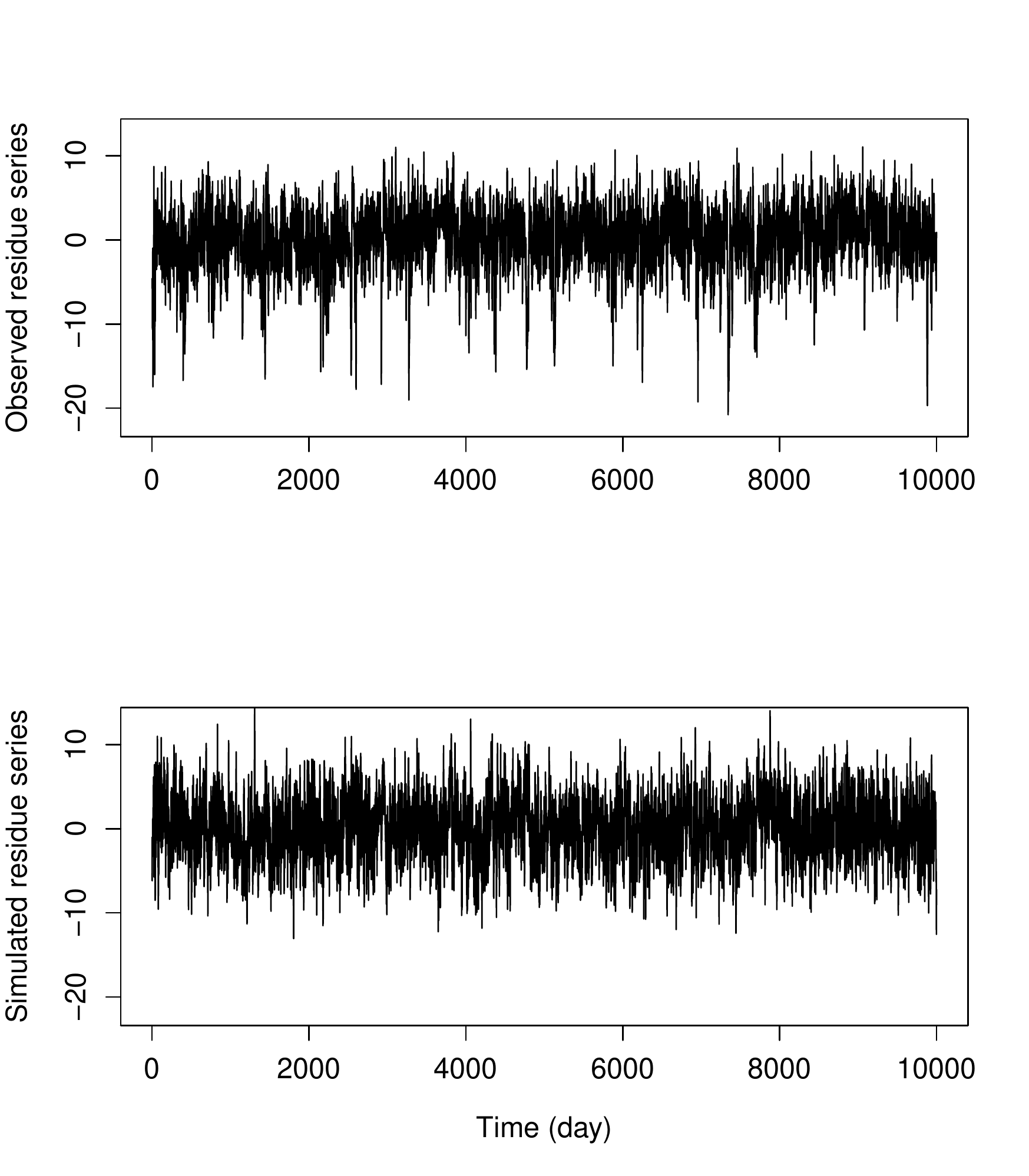}
    \caption{Time series plots of the observed seasonality-removed daily
    average temperature series and simulated seasonality-removed daily average
    temperature series. The magnitudes of variations in the observed
    seasonality-removed series do not match those in the simulated
    seasonality-removed series very well.}
    \label{fig:tempts}
\end{center}
\end{figure}

To see how well the daily average temperatures generated from the
fitted ARIMA model approximate the truth, we use diagnostic plots. Figure \ref{fig:acfpacf} shows the plots of the sample
auto-correlation function (ACF) and the partial ACF \citep{Chatfield2004} of
the seasonality-removed series and simulated seasonality-removed series. These
two series have similar correlation structures. However, the ACF and partial
ACF do not uniquely determine a time series. The time series plots of the two
series are shown in Figure \ref{fig:tempts}. We see that, the magnitudes of
variations in the seasonality-removed series are not symmetric about 0. At
some time points, the seasonality-removed series have exceptionally low
values, which we do not observe in the simulated seasonality-removed series.
The cause of this difference might be that we didn't account for the periodic
signals other than seasonal variation in the ARMIA model, or may be that the
noise in the seasonality-removed series are not inherently normal, issues to
be addressed in future work.

We now consider the prediction of bloom--dates. At the end of the current year, we generate 1000 series of the daily average temperatures of the whole next year
using the fitted ARIMA model. For each crop, we then use
(\ref{eq:predictivedMC}) to obtain the probability of blooming on each
successive day of the next year. This way we get a discrete predictive
distribution for bloom--date of the new year.

Now suppose that we are at the end of the first day of the new year with its
observed average daily temperature. We apply the fitted ARIMA model again to
generate 1000 series of temperatures starting from the second day of the new
year to the end of the year. We can then get another predictive distribution
for the bloom--date of the new year. We repeat this procedure on each
successive day, until the true bloom--date, at which time prediction ceases.
If the true bloom--date of the new year were day 120 for example, we would get
120 successive predictive distributions. What we expect to see are
increasingly more accurate predictions as the days progress toward the
bloom--date and more and more information about the daily averages
temperatures come to hand for that season. Growing confidence in that
prediction would provide an increasingly strong basis for management
decisions.

To see if our expectaions are realized, we perform a leave-one-out prediction
procedure -- at every step, for each crop, leave out one year of data for
prediction assessment and use the remaining years for training the model. Let's
consider apple as an example. For each left-out year, we follow the above
prediction scenario. We thus get 28 years (1937--1964) of assessments, with a
total of 3643 predictive distributions. For each of these predictive
distributions, we calculate the mean, median and mode as possible candidates
for point predictions of the new bloom--date. Also, we calculate a quantile based
95\% prediction interval (PI) for the new bloom--date. With all the 3643
predictive distributions, we can then estimate the root mean square error
(RMSE) and the mean abosolute errors (MAE) of the point predictions, and the
coverage probability and the average length of the 95\% PI. The results are
shown in Table \ref{tab:predscore}.
\begin{table}[!ht]
\begin{center}
  \caption{Cross validation results: The RMSEs and MAEs for point predictions
  using mode, median and mean are shown in column 2--7. The estimated
  coverages and average lengths of the 95\% PIs are shown in the last two
  column respectively. The units for RMSE, MAE and average length of the 95\%
  PI are day. The estimated coverage probabilities of these 95\% PIs are
  generally too high.}
  \label{tab:predscore}
  \begin{tabular}{|l|cc|cc|cc|cc|}
    \hline
    &\multicolumn{2}{c|}{Mode} &\multicolumn{2}{c|}{Median}
      &\multicolumn{2}{c|}{Mean} &\multicolumn{2}{c|}{95\% PI} \\
    \raisebox{1.0ex}[0cm][0cm]{Crop}
    & RMSE & MAE & RMSE & MAE & RMSE & MAE 
    & Coverage & Ave. Len. \\
    \hline
    Apricot & 6.74 & 5.12 & 6.58 & 5.06 & 6.62 & 5.14 & 0.99 & 33.24 \\
    Cherry  & 6.76 & 4.97 & 6.59 & 4.88 & 6.59 & 4.92 & 0.99 & 34.29  \\
    Peach   & 5.43 & 4.09 & 5.33 & 4.04 & 5.34 & 4.05 & 0.99 & 28.41 \\
    Prune   & 5.82 & 4.31 & 5.45 & 4.11 & 5.46 & 4.16 & 0.99 & 30.55 \\
    Pear   & 5.60 & 4.19 & 5.65 & 4.36 & 5.69 & 4.40 & 0.99 & 29.99 \\
    Apple   & 5.39 & 4.07 & 5.44 & 4.19 & 5.45 & 4.23 & 0.99 & 28.86 \\
    \hline
  \end{tabular}
\end{center}
\end{table}
\begin{table}[ht]
\begin{center}
  \caption{Maximum, minimum and range of the observed bloom--dates for each crop
  in 1937--1964 in the Okanagan region}
  \label{tab:datarange}
\begin{tabular}{|l|cccccc|}
  \hline
 & Apricot & Cherry & Peach & Prune & Pear & Apple \\ 
  \hline
  Maximum (day) & 126 & 136 & 135 & 138 & 139 & 146 \\ 
  Minimum (day) & 94 & 102 & 105 & 111 & 110 & 115 \\ 
  Range (day) & 32 & 34 & 30 & 27 & 29 & 31 \\ 
   \hline
\end{tabular}
\end{center}
\end{table}

We also provide in Table \ref{tab:datarange} the observed range of the bloom
dates (the difference time between the maximal and minimal observed bloom
dates) of each crop as a measure of natural variation of the bloom--dates for
each crop. The mean, median and mode as point predictions perform roughly the
same in terms RMSE and MAE. The RMSEs for all crops fall between 5.3 and 6.8
days, and the MAEs fall between 4.0 to 5.2 days. Considering the observed
ranges of the bloom--dates, which vary from 27 to 34, our point predictions
provide more useful information about the future bloom--dates. The estimated
coverage probabilities of 95\% PIs are too high for all crops, relative to 
the expected 95\%. For each crop, the average length of the 95\% PI
is roughly the same as the observed range of the bloom--dates, in accord with
the high estimated coverage probability.
These imply that our 95\% PIs
incorporate too much uncertainty,
possibly because that in the ARIMA model, we have included too much
variability caused by periodic signals other than seasonal variation as the
variability of the random noise.

We therefore reduce the variance of the white noise in the ARIMA model to half
the estimated value, while keeping all the other estimated parameter
unchanged.  We use this new ARIMA model to generate daily average
temperatures, and perform the above cross validation procedure again. The
results (Table \ref{tab:predscoreInnoP4}) show that while the accuracy of the
point predictions is roughly the same as before, the estimated coverage
probabilities and average lengths of the 95\% PIs are reduced to reasonable
values. This result does not confirm that the high estimated coverage
probabilities are actually caused by the high uncertainty in the predicted
daily average temperatures, but it at least adds weight to this explanation.
\begin{table}[!ht]
\begin{center}
  \caption{Cross validation results when using variance reduced simulated
  daily average temperatures: The RMSEs and MAEs for point predictions using
  mode, median and mean are shown in column 2--7. The estimated coverages and
  average lengths of the 95\% PIs are shown in the last two column
  respectively. The units for RMSE, MAE and average length of the 95\% PI are
  day. The estimated coverage probabilities of these 95\% PIs are reasonable.}
  \label{tab:predscoreInnoP4}
  \begin{tabular}{|l|cc|cc|cc|cc|}
    \hline
    &\multicolumn{2}{c|}{Mode} &\multicolumn{2}{c|}{Median}
      &\multicolumn{2}{c|}{Mean} &\multicolumn{2}{c|}{95\% PI} \\
    \raisebox{1.0ex}[0cm][0cm]{Crop}
    & RMSE & MAE & RMSE & MAE & RMSE & MAE 
    & Coverage & Ave. Len. \\
    \hline
    Apricot & 6.91 & 5.20 & 6.78 & 5.12 & 6.72 & 5.08 & 0.94 & 24.87 \\
    Cherry  & 6.62 & 5.06 & 6.64 & 5.05 & 6.58 & 4.99 & 0.93 & 26.46 \\
    Peach  & 5.56 & 4.03 & 5.51 & 3.95 & 5.49 & 3.98 & 0.95 & 21.17 \\
    Prune  & 5.48 & 4.16 & 5.46 & 4.04 & 5.46 & 4.07 & 0.98 & 22.38  \\
    Pear  & 5.98 & 4.46 & 5.79 & 4.33 & 5.75 & 4.31 & 0.94 & 21.45 \\
    Apple  & 5.76 & 4.36 & 5.53 & 4.17 & 5.48 & 4.15 & 0.95 & 20.36 \\
    \hline
  \end{tabular}
\end{center}
\end{table}

The above results for predictions derive from two models: Model AGDD for
blooming event and the ARIMA model for daily average temperature. To check the
performance of Model AGDD solely, we assume all the future daily average
temperatures are known, and then perform the above leave-one-out procedure
again. The results are reported in Table \ref{tab:predscoreknowntemp}. Note
that, in this case, since the uncertainty of the future daily average
temperatures is totally eliminated, we can only get one predictive
distribution for each test year. Therefore we cannot give a sensible estimate
for the coverage probability of the 95\% PIs. But we do see that the accuracies of
these point predictions are much higher than those of our previous
predictions, and the average lengths of the 95\% PIs are much smaller.
Although these are no longer real predictions, the results tend to validate
our Model AGDD for blooming event. Also, this finding shows the importance of
accurately modeling the covariate series and points to the need to improve the
temperature forecasting models.

\begin{table}[!ht]
\begin{center}
  \caption{Cross validation results if future daily average temperatures were
  known: The RMSEs and MAEs for point predictions using mode, median and
  mean are shown in column 2--7. The average lengths of the 95\% PI are shown
  in the last column. The units for RMSE, MAE and average length of the 95\%
  PI are day. The point predictions are very accurate, and the average lengths
  of the 95\% PIs are short.}
  \label{tab:predscoreknowntemp}
  \begin{tabular}{|l|cc|cc|cc|c|}
    \hline
    &\multicolumn{2}{c|}{Mode} &\multicolumn{2}{c|}{Median}
      &\multicolumn{2}{c|}{Mean} & 95\% PI \\
    \raisebox{1.0ex}[0cm][0cm]{Crop}
    & RMSE & MAE & RMSE & MAE & RMSE & MAE 
    & Average Length \\
    \hline
    Apricot & 4.18 & 3.30 & 3.65 & 2.93 & 3.62 & 2.90 & 13.48 \\ 
    Cherry  & 3.74 & 2.93 & 3.39 & 2.43 & 3.46 & 2.42 & 18.43 \\ 
    Peach   & 3.43 & 2.85 & 3.35 & 2.78 & 3.27 & 2.76 & 12.67 \\ 
    Prune   & 3.06 & 2.54 & 2.98 & 2.50 & 2.93 & 2.36 & 11.46 \\ 
    Pear    & 2.93 & 2.11 & 2.64 & 1.89 & 2.67 & 2.00 & 9.21 \\ 
    Apple   & 2.12 & 1.86 & 2.04 & 1.71 & 1.97 & 1.61 & 8.29 \\
    \hline
  \end{tabular}
\end{center}
\end{table}

\subsection{More about predictive uncertainties}

\begin{figure}[!ht]
\begin{center}
    \includegraphics[scale=0.6]{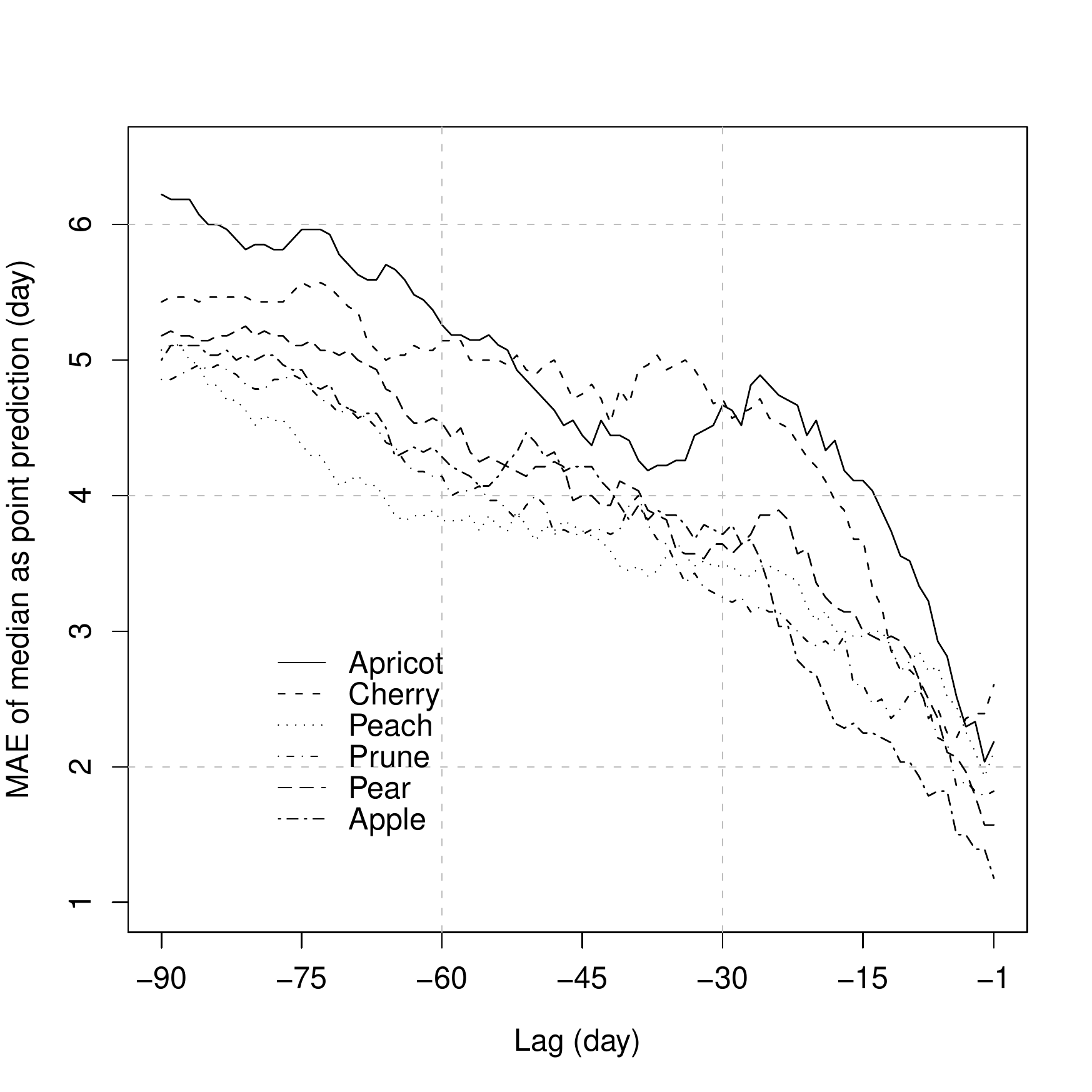}
    \caption{Change of the MAE of median with the change of lag. The point
    prediction becomes more accurate when time approaches the bloom
    date.}
    \label{fig:uncMAE}
\end{center}
\end{figure}

\begin{figure}[!ht]
\begin{center}
    \includegraphics[scale=0.6]{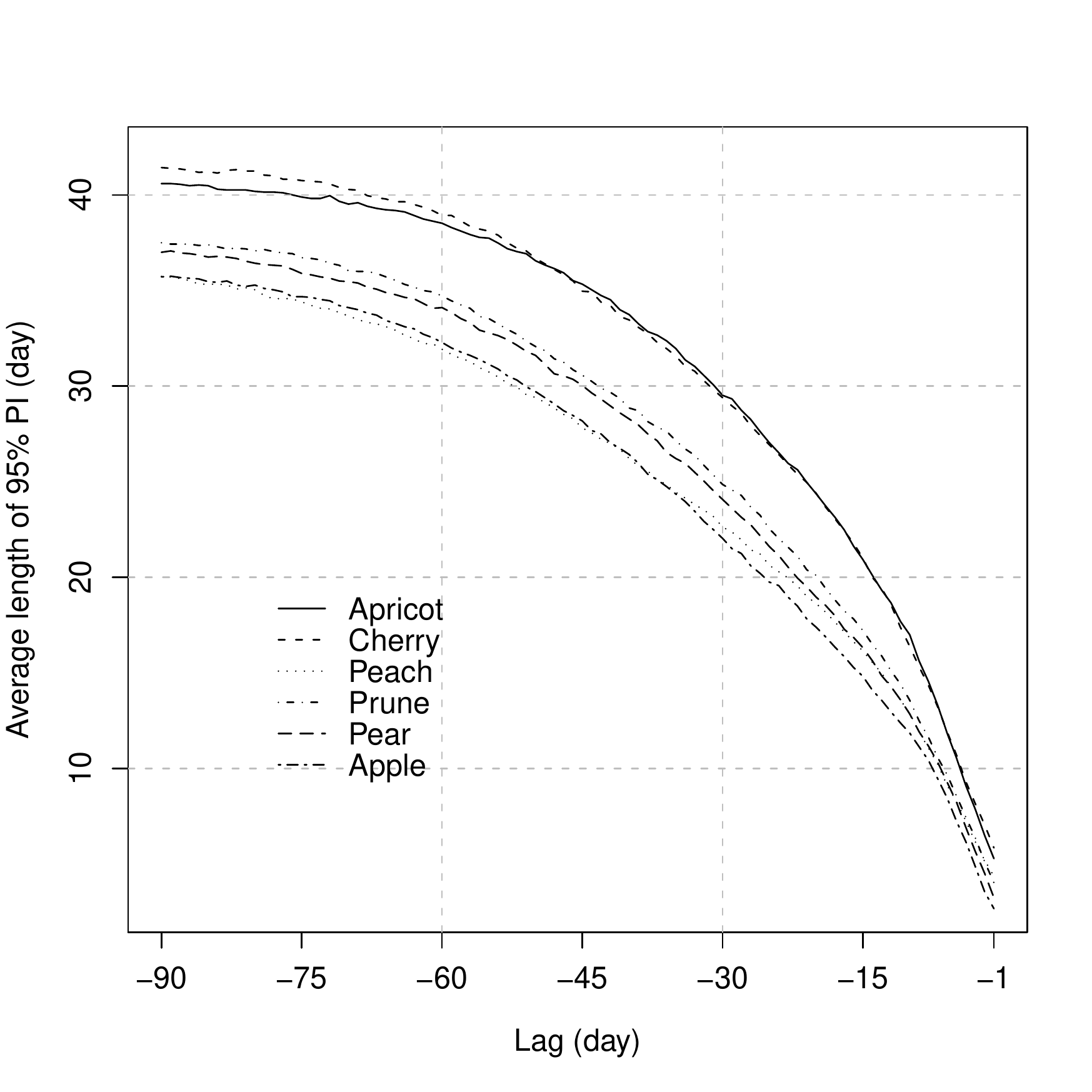}
    \caption{Change of the average length of 95\% PIs with the change of lag.
    The predictive uncertainty decreases when time approaches the bloom
    date.}
    \label{fig:uncPI}
\end{center}
\end{figure}

As noted above, we expected our point prediction to become more accurate and
predictive uncertainty to become smaller as time approaches the true bloom
date. To check this, for each crop, we calculate the MAE with median as point
prediction and the average lengths of 95\% PIs each day over the years of
interest, starting from 90 days prior to the bloom--date (call it "lag -90") to
1 day prior to the bloom--date ("lag -1"). The results are shown in Figure
\ref{fig:uncMAE} and \ref{fig:uncPI} respectively. It is clear that for all
crops, the MAE does become smaller and the average lenghth of 95\% PIs becomes
shorter as time approaches the true bloom--date.
In fact, by the time we reach
one month prior the bloom--date, the point prediction is quite accurate (the
MAE is 3.5--5 days).

\begin{figure}[!ht]
\begin{center}
    \includegraphics[scale=0.6]{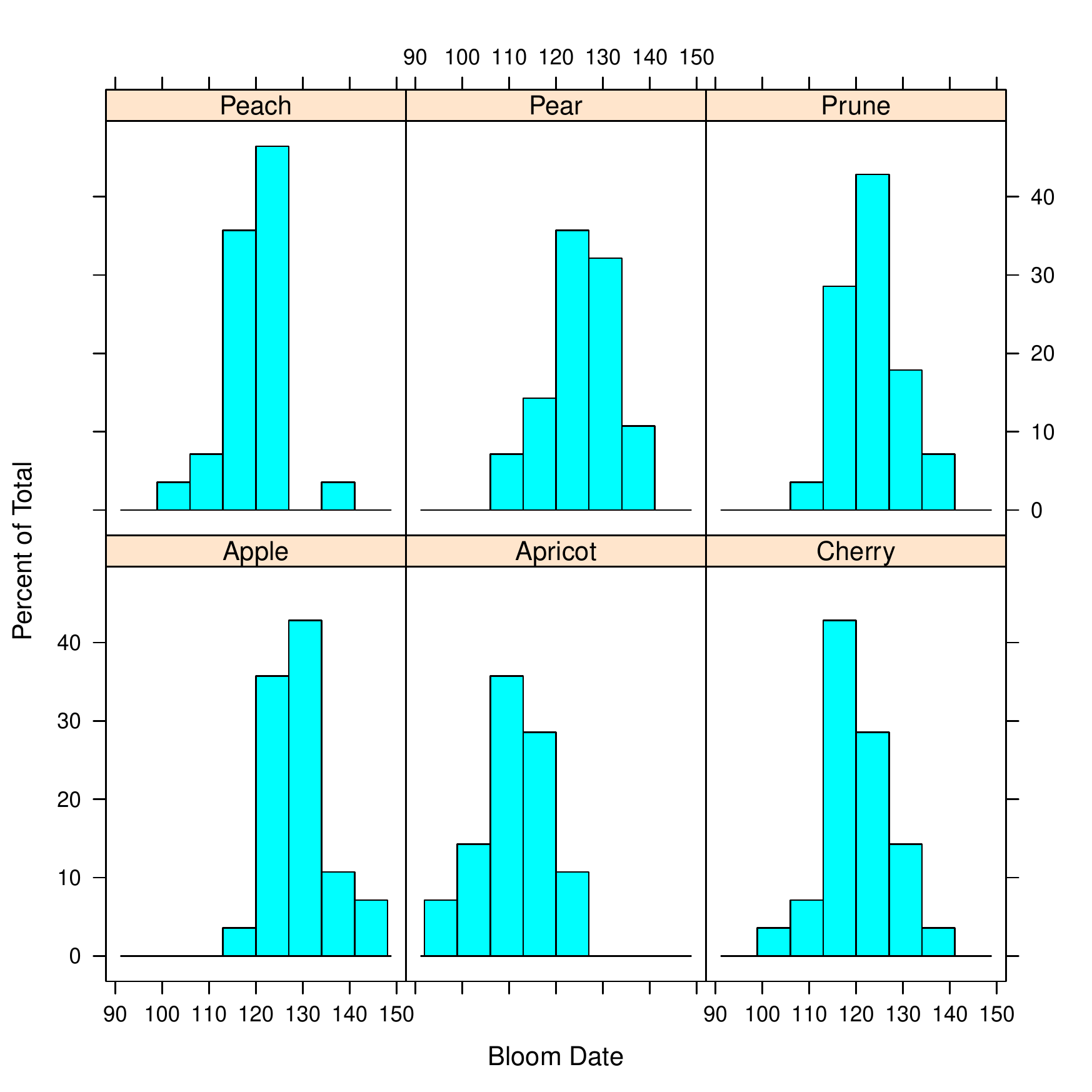}
    \caption{Histograms of the bloom dates of the corps. Most of the
    histograms are obviously skewed.}
    \label{fig:3histogram}
\end{center}
\end{figure}

We now compare our predictor with two naive predictors: the first one being
the 95\% confidence interval (CI) of a normal fit to the observed data, i.e.
$\pm1.96$ standard deviation around the sample mean; the second one being an
empirical quantile-based 95\% CI of the observed data, i.e. 2.5\%--97.5\%
sample quantiles. The length and percentage of the coverage of the normal fits
are shown in Table \ref{tab:PINormal}. Although the length of the normal CIs
are a few days shorter than our overall 95\% PIs reported in Table
\ref{tab:predscore}, the coverages of the normal CIs are too low, especially
for peach, prune and pear. Figure \ref{fig:3histogram} shows that the
histograms of the bloom dates of the crops are obviously skewed, which implies
that the normal fit may not be a good choice. The length and coverage of the
empirical quantile-based CIs are shown in Table \ref{tab:PIEmpirical}. These
CIs beat the Normal CIs in both length and coverage for most of the crops.
However, the coverages of them are still lower than 95\% except for Pear.
The results reported in Table \ref{tab:predscore} for our predictor are
calculated from the first day of a year to the
actual bloom date, which have incorporated too much uncertainty. If we look at
the predictions starting from one month prior to the bloom date to the actual
bloom date, the results (Table \ref{tab:PIonemonth}) are much improved -- the
average lengths of the 95\% PIs are much shorter and the estimated coverage
probabilities now range from 95\% to 98\%. These results apparently are much
better than those from both naive predictors.
\begin{table}[ht]
\begin{center}
  \caption{The length and the coverage of 95\% CI of a normal fit to the observed
  bloom--dates for each crop in 1937--1964 in the Okanagan region}
  \label{tab:PINormal}
\begin{tabular}{|l|cccccc|}
  \hline
  &Apricot & Cherry & Peach & Prune & Pear & Apple \\ 
  \hline
  Length &30.65 & 29.23 & 27.20 & 25.62 & 26.76 & 25.57 \\
  coverage &0.93 &0.93 &0.89 &0.89 &0.93 &0.93 \\
  \hline
\end{tabular}
\end{center}
\end{table}
\begin{table}[ht]
\begin{center}
  \caption{The length and the coverage of the empirical quantile-based 95\% CI of
  the observed bloom--dates for each crop in 1937--1964 in the Okanagan
  region}
  \label{tab:PIEmpirical}
\begin{tabular}{|l|cccccc|}
  \hline
  &Apricot & Cherry & Peach & Prune & Pear & Apple \\ 
  \hline
  Length &28.10 &29.28 &24.15 &24.97 &27.65 &24.25 \\
  coverage &0.93 &0.93 &0.93 &0.93 &0.96 &0.93 \\
  \hline
\end{tabular}
\end{center}
\end{table}
\begin{table}[ht]
\begin{center}
  \caption{The average length and the estimated coverage probability of 95\%
  PIs of our predictions starting from one month prior to the bloom date to
  the actual bloom date}
  \label{tab:PIonemonth}
\begin{tabular}{|l|cccccc|}
  \hline
  &Apricot & Cherry & Peach & Prune & Pear & Apple \\ 
  \hline
  Length   & 19.91 & 19.87 & 15.28 & 16.38 & 15.57 & 14.19 \\
  coverage & 0.95  & 0.96  & 0.97  & 0.98  & 0.97  & 0.97  \\
  \hline
\end{tabular}
\end{center}
\end{table}

\begin{figure}[!ht]
\begin{center}
    \includegraphics[scale=0.6]{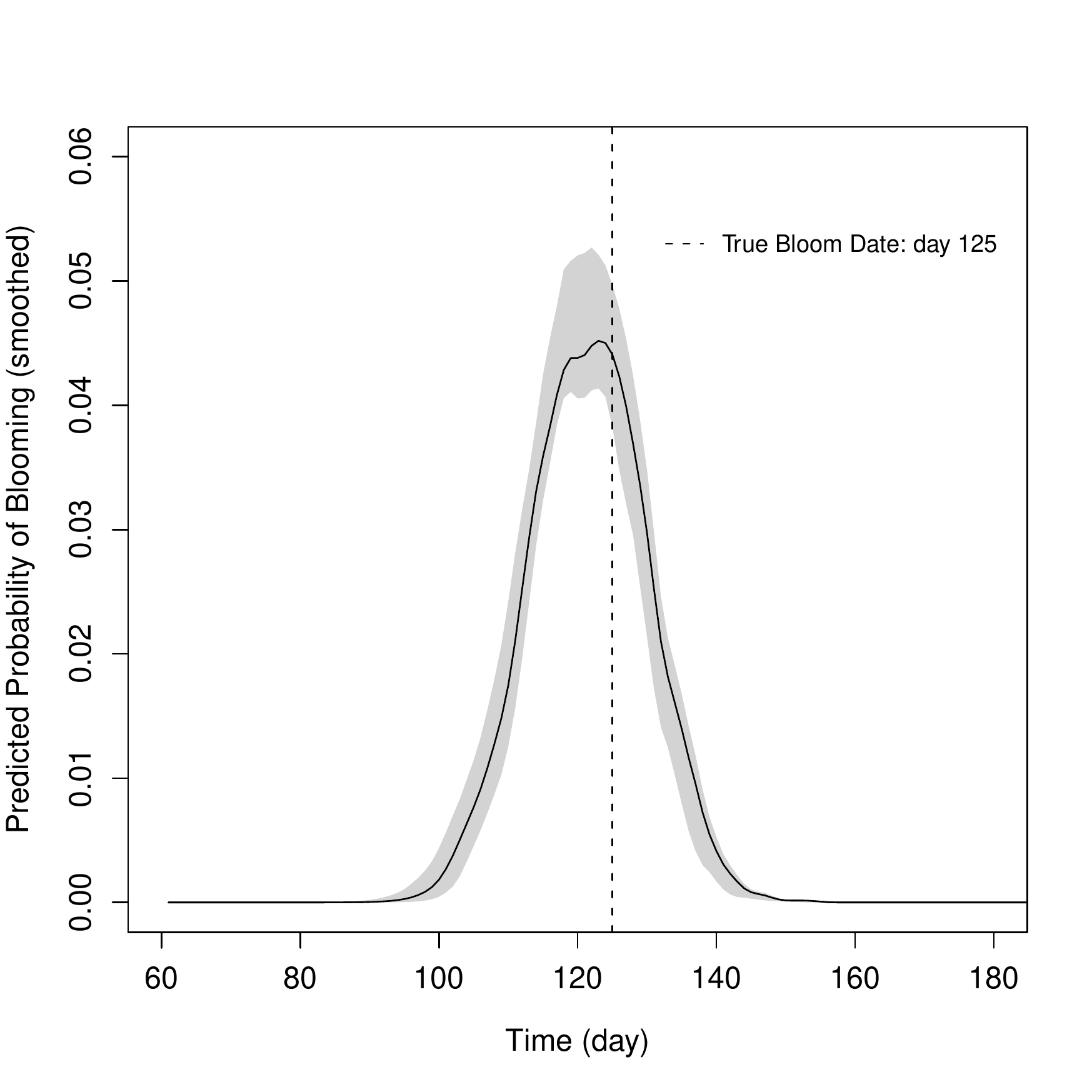}
    \caption{The predictive distribution (solid curve) of peach in year 1944
    with daily average temperatures of the first 60 days of that year known.
    The shaded area is a 95\% confidence band for this predictive
    distribution. The true bloom -- date of peach in that year is day 125.}
    \label{fig:pdplot}
\end{center}
\end{figure}

Another thing that interests us is the shape of the predictive distributions.
To see this, we plot the predictive distribution of a ``randomly'' picked crop
and test year -- the predictive distribution of peach in year 1944 with daily
average temperatures of the first 60 days of that year known (Figure
\ref{fig:pdplot}). Note that the true bloom--date of peach in that year is day
125, and we have smoothed the discrete predictive distribution to a continuous
curve. We see that the predictive distribution (the solid curve) is nearly
bell-shaped which roughly looks like a normal distribution. Since this
predictive distribution is calculated by plugging in the MLEs as if they were
the true parameters, there is an uncertainty associated with this predictive
distribution. Just as before, we use the bootstrap to assess this
uncertainty -- we calculate a quantile based 95\% bootstrap confidence band
(shown as the shaded area in Figure \ref{fig:pdplot}) for this predictive
distribution. The plot shows that this confidence band is not too wide, so we
basically can ``trust'' this predictive distribution.

The validity of the bootstrap procedure arises again.
Do the bootstrap estimates of the quantiles of the predictive probabilities
reflect the true quantiles of the predictive probabilities? Again, we conduct
a simulation study to answer this question. Take the same settings for the
simulation as described in Section \ref{subsec:est}. For each sample size $S$,
where $S \in \left\{ 30, \,80, \,150, \,400\right\}$, we now generate one more
year of data as test data. For a fixed sample size $S$, for each sample, we
estimate the model parameters, and we then use this set of of parameters to
predict the bloom--date of the test year by assuming the first 60 days of
temperatures are known. We then get 1000 predictive distributions for each
sample size. For each future day, we take the 2.5\% and 97.5\% sample
quantiles of the 1000 predictive probabilities to approximate a quantile based
95\% confidence interval for the predictive probability. Now, randomly pick
one sample, and then take 1000 bootstrap samples of this sample, and estimate
model parameters using each bootstrap sample. With each set of estimated
parameters obtained from the bootstrap, we can then make a prediction on the
test year. With 1000 bootstrap samples, we get 1000 predictive distributions.
For each future day, as with the simulated data, we can obtain a quantile
based 95\% bootstrap confidence interval for the predictive probability. We
now compare the confidence intervals obtained in these two ways. Randomly
picking one future day, the 95\% confidence intervals for the predictive
probability of blooming that day calculated using the simulated data and
bootstrap are shown in Table \ref{tab:predCIsim}. We can see that both types
of confidence intervals become narrower when sample size becomes larger. For
each sample size, the bootstrap interval is close to the interval obtained
using the simulated data. Moreover, when the sample size reaches 400, the two
types of intervals are basically identical. This result suggests that applying
bootstrap might be a reasonable way to estimate the uncertainty of the
predictive probabilities.
\begin{table}[ht]
  \caption{Comparison of the 95\% confidence intervals for predictive
  probabilities obtained using bootstrap and the simulated data.}
  \label{tab:predCIsim}
\begin{center}
  \small
\begin{tabular}{|l|c|c|c|c|}
  \hline
  & $S=30$ & $S=80$ & $S=150$ & $S=400$ \\
  \hline
Bootstrap & (0.0300, 0.0350) & (0.0321, 0.0358) & (0.0300, 0.0320)
  & (0.0305, 0.0322) \\ 
  \hline
Simulation & (0.0288, 0.0348) & (0.0298, 0.0332) & (0.0302, 0.0326)
  & (0.0307, 0.0320) \\ 
   \hline
\end{tabular}
\end{center}
\end{table}

\section{Conclusions}\label{sec:conclusions}

In this paper, we presented an application of our regression model for
irreversible progressive events to a single phenological event -- blooming
event of six high-valued perennial crops. The approach we introduced here can
be also widely applicable to the analysis of survival data and other similar
time-to-event data.

For the blooming events of the six crops, our method provides a sensible way
to estimate the important parameter $T_{base}$ in the definition of growing degree day (GDD) and
our statistical analysis supports an earlier empirical finding -- that the timing
of a bloom event is related to AGDD, a cumulative unweighted sum of GDDs. Our method also provides useful point predictions of the future bloom--dates, as well as the assessment of the predictive uncertainties which is useful for risk
analysts and policy makers. Our 95\% PIs are excessively large however,
quite probably because we have used a crude ARIMA model for predicting the future daily average temperatures. After reducing and then total eliminating the
uncertainty about the daily average temperatures, we see increased accuracy of
point predictions and much shortened 95\% PIs. This observation validates our
regression model for phenological events.

In our analysis, we did not consider the possible auto-correlation of the bloom
date. In future, we will consider more complicated models to incorporate
this. On the other hand, to improve estimation and prediction, we may build a
multivariate model to encompass all crops as responses simultaneously.
Moreover, if crops of multiple locations are involved, we will consider a
spatial-temporal modeling approach for phenological events.

\afterpage{\clearpage}


\bibliographystyle{plainnat}
\bibliography{phenology_arXiv.bib}

\end{document}